\documentclass{emulateapj}

\def\etal{\it et al. \rm }
\begin{document}

\title{The Ages of Dwarf Ellipticals}

\author{Karl Rakos}
\affil{Institute for Astronomy, University of Vienna, A-1180, Wien, Austria;
karl.rakos@chello.at}

\author{James Schombert}
\affil{Department of Physics, University of Oregon, Eugene, OR 97403;
js@abyss.uoregon.edu}

\begin{abstract}

We present narrow band photometry of 91 dwarf ellipticals in the Coma and
Fornax clusters taken through the Str\"omgren ($uvby$) filter system.  Dividing
the sample by dwarf morphology into nucleated (dEN) and non-nucleated (dE)
dwarfs reveals two distinct populations of early-type systems based on
integrated colors.  The class of dEN galaxies are redder in their continuum
colors as compared to bright cluster ellipticals and dE type dwarfs, and
their position in multi-color diagrams can only be explained by an older mean
age for their underlying stellar populations.  By comparison with the narrow band
photometry of the M87 globular cluster system (Jordan \etal 2002), we find
that dENs are a higher metallicity continuation of the old, metal-poor color
sequence of galactic globulars and the blue population of M87 globulars.
Bright ellipticals and dE dwarfs, on the other hand, follow the color sequence
of the metal-rich, red population of M87 globulars.  A comparison to SED models,
convolved to a simple metallicity model, finds that dENs and blue globulars
are 3 to 4 Gyrs older than cluster ellipticals and 5 Gyrs older than dE type
galaxies.  The implication is that globulars and dEN galaxies are primordial
and have metallicities set by external constraints such as the enrichment of
their formation clouds.  Bright ellipticals and dE galaxies have metallicities
and ages that suggest an extended phase of initial star formation to produce
a younger mean age, even if their formation epoch is similar to that of dENs
and blue globulars, and an internally driven chemical evolutionary history.

\end{abstract}


\section{INTRODUCTION}

The study of stellar populations in galaxies has focused on ellipticals,
due to their uniform properties as a class of objects. Early work proposed
a simple formation scenario (the so-called monolithic collapse model,
Eggen, Lynden-Bell \& Sandage 1962, Larson 1974, Arimoto \& Yoshii 1987).
However, ellipticals span a large range of masses and the dwarf
ellipticals and, while they are similar in morphology,
may follow very different evolutionary paths.  Foremost in
the differences between dwarf and bright ellipticals is the hypothesis
that dwarf ellipticals are primordial and that large galaxies are
constructed from the merging of dwarf galaxies (hierarchical models,
Davis \etal 1985, White \& Frenk 1991, Kauffmann \& Charlot 1998,
Somerville \& Primack 1999).  Thus, the age of stellar populations in
dwarf ellipticals can be a key test of our galaxy formation scenarios.

The uniformity of bright ellipticals in terms of age and metallicity has
been demonstrated in various ways.  For example, the low scatter in the
color-magnitude relation for ellipticals argues for a very limited range
in age at a particular luminosity (Terlevich \etal 1999).  While there is
evidence for an intermediate age population in some ellipticals (Trager
\etal 2000), spectroscopic studies have convincingly demonstrated that a
majority of bright cluster ellipticals are old and the color-magnitude
relation is dominated by metallicity effects (Terlevich \& Forbes 2002,
Kuntschner 2000).  A uniform spread of metallicity in ellipticals is
consistent with the color-[Fe/H] relation (Rakos \etal 2001) which implies
an early, and rapid, phase of star formation.

The age and chemical evolution of dwarf ellipticals is less well known, mostly
due to their fainter luminosities and lower central surface brightnesses.
Pre-21st century results on the stellar populations in dwarf ellipticals
focused on their optical/near-IR broadband colors with a small sample of
spectroscopic results for objects in Virgo and Fornax (see Ferguson \& Binggeli
1994 for a review on the colors and spectra of dEs).  The broadband color
results were ambiguous with respect to age, with most studies noting that dE
colors were more similar to galactic globular colors than to cluster ellipticals.
Near-IR colors suggest that dEs behave as metal-rich globulars in the
$B-H$,$J-K$ diagram, but that some dEs have bluer $B-H$ colors (Bothun \etal
1985) suggestive of a young population.  Spectroscopy of small dE samples
(Bothun \& Mould 1988, Held \& Mould 1994) found that nucleated dwarfs (dEN)
basically follow the same locus of spectral indices (such as H$\beta$) as
galactic globulars.  There were several notable exceptions of dwarfs with
strong Balmer lines which indicated a young mean age or an intermediate age
population, and this agrees with the very recent observations of Caldwell
\etal (2003) who find the scatter in Balmer line strengths to increase with
lower velocity dispersion (lower mass) systems.

With respect to stellar populations of dEs versus dENs, only the Caldwell
\& Bothun (1987) study performed a detail color study of the two types in
the Fornax cluster.  They found a color difference between nucleated dwarfs
in the sense that dENs with brighter nuclei tended to have redder $U-B$
integrated colors, but found no difference between the color of the nucleus
and the colors of the stellar envelope that makes up most of the galaxy's
light.  This was an unexpected result since the simplest explanation for
the nucleation in dEs is a recent episode of star formation to produce a
central cluster.  Such an event should produce a different color for the
nucleus compared with the envelope of the dE, but this is not observed in
any sample.  The trend of redder nucleus (and galaxy) color with
luminosity of the nucleus was interpreted by Caldwell \& Bothun as
reflecting the mass-metallicity relation seen in larger galaxies.
Caldwell \& Bothun proposed a scenario where the brighter (i.e.  more
massive) nuclei have strong gravitational fields which not only hold onto
more gas to produce more stars, but also more metals to produce redder
colors.  Presumingly, this occurred coeval with the rest of the dwarf
galaxy's star formation such that the color difference due to age is
undetected.

The purpose of this project is to examine the narrow band Str\"omgren colors
($uvby$) for a large sample of dE and dENs in Fornax and Coma.  Our objectives
are threefold: 1) to make a direct comparison of dwarf and bright elliptical
colors and test for any property, from their colors, that signal differences
in their underlying stellar populations (raw empirical tests), 2) compare
their colors to colors of galactic and extragalactic globular clusters, the
simplest stellar populations, (interpretation guided only by data) and 3)
compare their colors to predictions from various spectral energy distribution
(SED) models (speculation guided by stellar physics).  There is no direct
method to determine the age of stellar populations in galaxies through colors
or spectroscopy, so we are forced to use SED models to quantify our numbers.
However, the relative ages between galaxy types is independent of models and
can answer several key questions to the early stages of galaxy evolution.

\section{OBSERVATIONS}

\subsection{M87 Globular Cluster System}

In our previous work on the narrow band colors of elliptical galaxies,
it was useful to make a comparison with simple stellar populations (SSP), i.e.
globular clusters.  To this end, we have established a metallicity and age
baseline using globular clusters from the SMC, LMC and Milky Way (see Rakos
\etal 2001).  While this was sufficient to provide calibration [Fe/H] values
for our multi-metallicity models, the globular cluster data lacked sufficient
age range to resolve mean age issues and, thus, SED models were consulted.

Narrow band data are sparse for globulars clusters, however a new HST narrow
band study of the globular cluster system around M87 was recently published
(Jordan \etal 2002).  These data were obtained in filters F336W, F410M, F467M
and F547M using the Wide Field Planetary Camera 2 (WFPC2) and closely match
the Str\"omgren filters used for our study (although the F336W filter is
slightly wider than the Str\"omgren $u$ filter).  The M87 photometry was
calibrated to the VEGAMAG system (HST Handbook) which we have converted to
our $uvby$ system using values of $uz-vz=+0.937$, $bz-yz=-0.218$ and
$vz-yz=-0.664$ for Vega.  For our analysis, we have trimmed the Jordan
\etal sample to those objects brighter than 23.5 $m_{5500}$.  This was
done primarily because the photometric errors sharply increase for mags
fainter than 23.5; however, we also wish to compare the M87 globulars to
our Milky Way sample and the faintest galactic globular in our sample is
approximately $M_V=-7$, which corresponds to $m_{5500}=23.5$ for a
distance modulus of 31.43 to M87 (Tonry \etal 2001).  The final trimmed
sample contained 443 objects.

\begin{figure*}
   \mbox{}\hfill
   \begin{minipage}[t]{0.99\linewidth}
     \centerline{\includegraphics[width=0.48\linewidth]{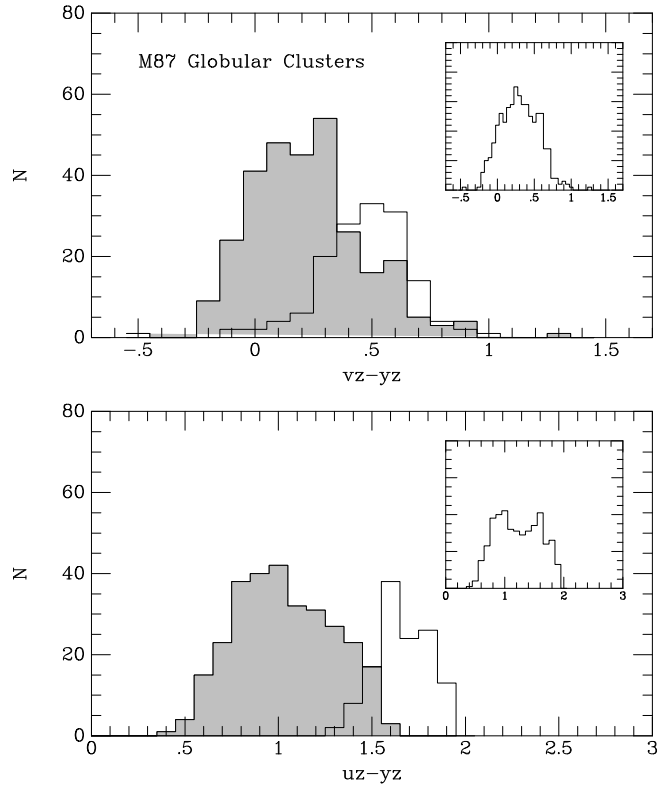}}
\caption{Histograms of $uz-yz$ and $vz-yz$ color for the M87 globular
cluster population from Jordan \etal (all objects brighter than $m_{5500}=23.5$).
The two samples are selected by PC analysis as discussed in text, the shaded
sample has PC1 values less than one, the open sample has PC1 values greater
than one.  The bimodality to the entire sample is visible in both colors,
where the metal-poor population is the most numerous.  Inset histograms
display the entire sample.  }
   \end{minipage}
\end{figure*} 

The M87 globular cluster colors are presented as histograms in Figure 1.
We confirm the bimodality found by Jordan \etal in both $uz-yz$ and
$vz-yz$.  There are clear peaks at $uz-yz=0.9$ and $uz-yz=1.6$. The
bimodality is less obvious in $vz-yz$, but peaks at $vz-yz=0.2$ and
$vz-yz=0.6$ are visible.  There is no evidence for a third peak which
would correspond to a metal-rich, intermediate age cluster population
(Puzia \etal 2002).  Most extragalactic globular studies use $V-I$
photometry calibrated to [Fe/H] using galactic globulars.  While this
calibration is less than perfect, it does have the advantage of being
relatively insensitive to age effects.  The current $V-I$ to [Fe/H]
conversion estimates that the peak of the blue population occurs at $-$1.3
dex and the peak of the red at $-$0.2 dex.  However, in Rakos \etal
(2001), we directly calibrate $vz-yz$ to [Fe/H] using the same galactic
globular system, with a smaller scatter than $V-I$.  Comparison to SED
models confirms that this conversion should be accurate to 0.2 dex for
populations with mean ages greater than 8 Gyrs.  With this calibration,
the two peaks in the Jordan \etal data correspond to [Fe/H] = $-$1.7 and
$-$0.9 dex, about 0.5 dex more metal poor than the $V-I$ calibration.
Note that this difference in [Fe/H] values to those quoted by Jordan \etal
or Kundu \etal (1999) are due to differences in the $V-I$ calibration
versus $vz-yz$, not due to the use of different SED model assumptions.

Separate metallicity peaks suggests these are two distinct populations, thus
we can use principal component (PC) analysis to separate out the populations
in multi-color diagrams as we have successfully done on cluster populations
(Steindling, Brosch \& Rakos 2001).  Briefly, a principal component analysis,
in any n-dimensional space, calculates the axis along which the data points
present the largest, most significant scatter. This is called the first
principal component (PC1). It then proceeds to calculate PC2, the axis of
the second most significant spread in the remaining n-1 dimensional space
orthogonal to the first principal component, and so on.  Mathematically, this
is done by normalizing the coordinates of the data points to standardized
variables and calculating their covariance. The final output are the eigenvectors
(the principal components) and eigenvalues of this covariance matrix.  The
gain of the PC analysis is therefore twofold: (1) it minimizes the dimensionality
of the data and (2) it provides an orthonormal coordinate system in which
the data are most easily characterized.

To simplify comparison between the globular cluster populations and
elliptical galaxies, we have applied the same PC axes in our previous
papers, derived from all three narrow band colors in Coma ellipticals.
The boundary for ellipticals selected in Odell, Rakos \& Schombert (2002)
is the condition that PC1 be greater than one.  This same criterion very
nicely divides the M87 globulars into two distinct red and blue
populations with the same peaks as visible in the integrated population
(see Figure 1).  For the 443 clusters analyzed, 33\% are members of the
red population, 67\% are members of the blue.  A little numerical
experimentation finds that boundary values between 0.8 and 1.2 for PC1
produce similar distributions and, thus, our conclusions are insensitive
to the choice of PC axis values.

Figure 2 displays the multi-color diagrams for the M87 globulars and the
Milky Way globulars data from Rakos \etal (2001).  Here the M87 data have
been binned in constant $bz-yz$ bins for clarity with each bin containing
between 30 and 50 objects.  The $uz-yz$ data have been shifted 0.3 mags to
the red in what is an apparent mismatch in the VEGAMAG calibration to
$uvby$ values for Vega quoted above.  The source of this additive error is
not known (although is probably due to the wider F336W filter compared to
the normal Str\"omgren $u$), but is not critical to our discussions below
since the $uz-yz$ data are not a major component of our analysis.

The blue population (PC1 $<$ 1) has identical properties to the galactic
globulars, tracking the same range of metallicity where 90\% of the
globulars lies between $-$2.5 to $-$0.7.  The red population (PC1 $>$ 1)
lies in the region redder in $vz-yz$ for its $bz-yz$ colors than galactic
globulars.  Likewise, the red population has redder $uz-yz$ colors
compared to the blue population for a given $bz-yz$ value.
The separation of the two globular populations in color space is a
critical distinction since, if the difference was a simple one of metallicity,
the two populations would overlap in the $vz-yz$,$bz-yz$ diagram.  There
are many possible star formation histories that would result in values
that occupy this region of the two color diagram, however a younger age is
the most plausible reason to produce the $vz-yz$,$bz-yz$ trend since a
change solely in metallicity would track along the blue population as
demonstrated by the galactic globulars.

The common interpretation of the dual globular population in galaxies is
that the blue population is primordial and the red population was added at
a later date either by star formation or mergers with other galaxies (see
Kissler-Patig 2001).  Clearly, an age difference is a distinguishing
factor in the formation scenarios.  Our range of metallicity and
population numbers agree with the photometric studies, when adjusted for
the $V-I$ calibration difference (Gebhardt \& Kissler-Patig 1999).   Age
differences are best determined through the Balmer lines (a known age
indicator in old populations), which requires spectroscopy.  For example,
Cohen, Blakeslee \& Ryzhov (1998) examined 150 M87 globulars with high
resolution Keck spectra.  They find no difference in the two populations
with respect to Balmer lines at about the 2 Gyr sensitivity level.
However, the colors in Figure 2 are inconsistent with a single age for
both red and blue populations.  SED models are of little help in
quantifying the suspected age difference since they have an improper slope
at the low metallicity end of the globular color sequence.  The solid line
in Figure 2 is the 14 Gyr G\"ottingen SSP models (Schulz \etal 2002).
While a fair match for metallicities above $-$1.5, the galactic globulars
with [Fe/H] $< -2$ are much bluer in both $bz-yz$ and $vz-yz$ than
predicted by the models.  While there are numerous possible reasons for
this discrepancy (for example, a poor treatment of metal-poor HB stars in
the codes), it is beyond this study to investigate them all.  Despite the
lack of guidance from the SED models, the color difference between the two
globular populations in M87 is clear, and metallicity differences will not
account for the separation.  Thus, we are left with hotter continuum
colors for a given metallicity indicating an age difference.

\begin{figure*}
   \mbox{}\hfill
   \begin{minipage}[t]{0.99\linewidth}
     \centerline{\includegraphics[width=0.85\linewidth]{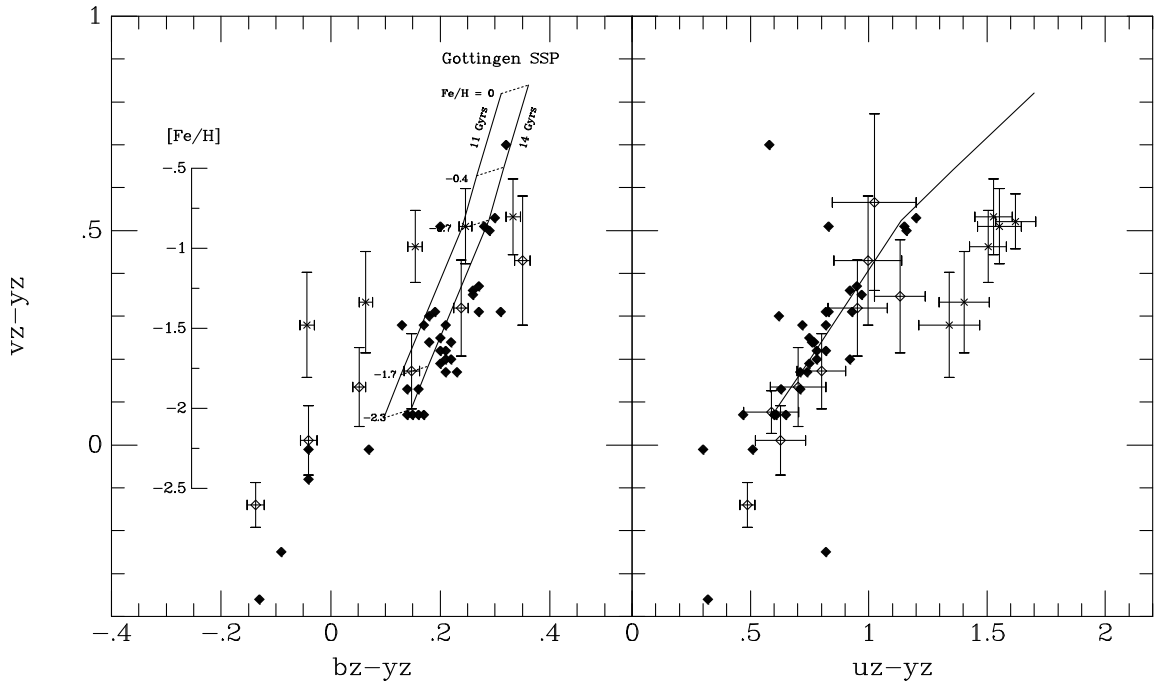}}
\caption{Two color diagrams for the M87 globular cluster population.  The
open symbols are the blue, metal-poor clusters, the crossed symbols are
the red, metal-rich clusters.  The data are binned in units of 40 clusters
per $bz-yz$ bin.  Galactic globulars from Rakos \etal (2001) are shown as
solid symbols.  The [Fe/H] calibration from Rakos \etal (2001) is shown as
an inset axis. For comparison, the 14 and 11 Gyr SSP models from Schulz
\etal 2002 are shown as solid lines, connected by dotted lines of constant
metallicity. There is no change in $uz-yz$ for a passive SSP over the 11
to 14 Gyr interval, thus a single model is shown in the right panel. The
blue population of M87 follows the Galactic globular data, but the red
population is significantly different from either group.} 
   \end{minipage}
\end{figure*}

Of interest to this paper is which of these two populations has the most in
common, in terms of mean colors, with the properties of dwarf and bright
ellipticals.  If the scenario where the red population is formed later is
correct, then its colors would match the colors of envelope stars rather
than galactic globulars.  If dwarf galaxies are older than bright
ellipticals, then we would expect their colors to match the colors of old,
metal-poor galactic globulars, albeit at higher metallicities due to the
mass-metallicity effect.  In any case, a comparison will be left until the
next section.

\subsection{Dwarf Ellipticals}

The multi-color data for the dwarf ellipticals in this sample derive
primarily from two previous studies in Fornax (Rakos \etal 2001) and Coma
(Odell, Rakos \& Schombert 2002).  The Fornax sample was selected from the
catalog of Ferguson (1989) and we have adopted his morphological
classifications.  The Coma sample was classified by visual inspection of
the $yz$ (5500\AA) CCD frames, guided by the appearance of the Fornax
dwarf ellipticals to maintain the same general scheme.  In addition, we
have placed an upper limit of $M_{5500}=-$19.5 ($-$18.5 in the blue) to
the sample, regardless of morphological classification, as the canonical
luminosity cutoff between giant and dwarf ellipticals.  For the Coma
sample, an additional six dwarf ellipticals were recovered off of frames
taken from the same telescope set-up as Odell, Rakos \& Schombert (2002),
but during a later observing run.  All the data were reduced in a like
manner as described in Rakos \etal (2001).  The final sample contained 9
dEs and 40 dENs in Fornax and 10 dEs and 32 dENs in Coma.  In addition, we
have used a distance modulus of 31.35 for Fornax (Madore \etal 1999) and
34.82 for Coma.

In our previous work on dwarf and bright cluster ellipticals (Odell, Rakos
\& Schombert 2002, Rakos \etal 2001) we treated the sequence from giant to
dwarf as a continuum, mostly due to the uniform nature of the color-magnitude
relation (CMR) and the smooth morphology of ellipticals of all luminosities.
For low redshift samples, the CMR has a consistent slope and zeropoint
from cluster to cluster, in agreement with a mass-metallicity
interpretation of the relationship.  However, as can be seen in Figure 3 of
Odell, Rakos \& Schombert (2002), the scatter increases significantly
below $M_{5500}=-17$.  This has been noted in numerous studies of the
broadband colors of dwarf ellipticals (see Bender \etal 1993) and Conselice,
Gallagher \& Wyse (2002) demonstrate that this scatter is a fundamental
breakdown of the linear relationship found in brighter ellipticals.
Possible interpretations for the increased scatter have focused on the chaotic
nature of star formation in low mass systems (such as variable rates of
baryonic blowout) and hierarchical mergers in order to produce a range of
chemical evolution histories.  Certainly intermediate age stars are evident
in nearby dEs (e.g. NGC 204) and it is entirely possible, in fact quite
likely, that age effects become increasing important for the colors of low
mass dwarf galaxies (see Caldwell \etal 2003).

\begin{figure*}
   \mbox{}\hfill
   \begin{minipage}[t]{0.99\linewidth}
     \centerline{\includegraphics[width=0.85\linewidth]{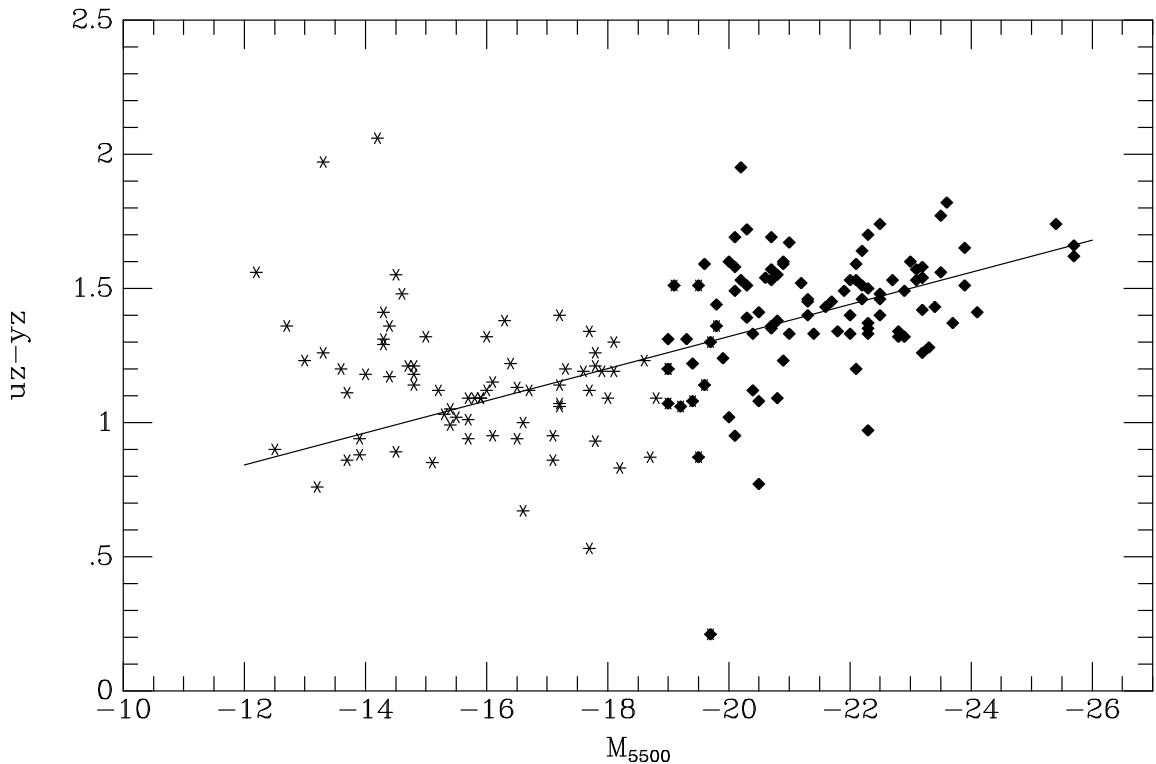}}
\caption{The near-UV color-magnitude diagram, $uz-yz$ versus absolute luminosity
$M_{5500}$ for Coma ellipticals, dwarf ellipticals and Fornax dwarf ellipticals.
The solid symbols are bright cluster ellipticals and crossed symbols are
dwarf ellipticals.  The traditional CMR is obvious for galaxies brighter than
$-$19 and argues for a self-enrichment star formation history through feedback
by supernovae.   The dwarf ellipticals, on the other hand, display no
correlation with luminosity (mass) which implies a different or chaotic
formation process.  The solid line is a least squares fit to the cluster
elliptical data, extrapolated to dwarf luminosities}
   \end{minipage}
\end{figure*} 

Given the history of the color-magnitude relation and its importance to galaxy
population studies, it is worthwhile revisiting the color-magnitude relation
in narrow band colors, particularly over the range of luminosity now available
with this new and larger sample.  Traditionally, the color-magnitude diagram
is plotted in $U-V$ to maximize the metallicity effect as seen across the
4000\AA\ break.  In Odell, Rakos \& Schombert (2002), we used the $vz-yz$
and $bz-yz$ indices since they are most sensitive to metallicity and age in
our narrow band system ($uz-yz$ is most sensitive to recent star formation
in our distant cluster work).  To provide a different perspective, in Figure
3 we present the $uz-yz$ colors (approximately $U-V$, albeit with narrow band
filters) for the bright ($M_{5500}<-19$) and dwarf ellipticals in Coma.  In
the past, $uz-yz$ has been a difficult indice to interpret for ellipticals.
Few of the early SED models made accurate predictions for the UV colors in old
populations. And the $uz-yz$ index is very sensitive to recent star formation or the
addition of young stars by a recent merger of a gas-rich spiral.  In any
case, our results are the same if $vz-yz$ or $bz-yz$ colors are used instead
of $uz-yz$.  The $uz-yz$ CMR for bright ellipticals is very clear, as is the
failure of the dwarfs to continue the same slope.  While the bright cluster
ellipticals have a strong correlation ($R$=0.43) taken by themselves, the
dwarf ellipticals display no correlation with luminosity and, hence, mass.
Since dwarf ellipticals do have lower metallicities than bright ellipticals,
based on spectroscopy (Held \& Mould 1994), this breakdown of the CMR signals
a real difference in the star formation history of dwarfs as compared to
massive cluster ellipticals.

Our first attempt to understand the lack of correlation in the CMR at low
luminosities involved dividing the dwarf elliptical sample by some other physical
characteristic, in this case, structural properties as given by morphology.
With regard to physical morphology, dwarf galaxies are found in a variety of
types and isophotal shapes.  Dwarf ellipticals are classified primarily by
the smooth elliptical shape of their isophotes and indirectly on their
gas-poor appearance (the presence of neutral gas being sharply correlated
with a lumpy luminosity distribution).  Within the dwarf elliptical
sequence there are dwarf ellipticals with distinct nuclei (dEN) and those
with smooth, lower surface brightness cores (dE).  Recent imaging has
found the nuclei of dENs to be similar in size and luminosity (mass) to
the largest galactic globulars. Proposed origin scenarios for nucleated
dEs include accretion of remnant cores of stripped galaxies (Gerola,
Carnevali \& Salpeter 1983), gas infall and star formation (Caldwell \&
Bothun 1987) or coalescence of several star clusters to form a single,
super-massive cluster (Oh \& Lin 2000).  Aside from the existence of a
nucleus, dE and dENs have similar properties with the expectation that, on
average, dEs are slightly flatter than dENs and dENs are more numerous
than dEs (Ferguson 1989).

The different cluster distances and telescope set-ups result in complimentary
samples for Fornax and Coma.  The Fornax sample reaches to deeper absolute
luminosities than the Coma sample ($-$13 versus $-$17) but, due to limited
areal coverage, has very few objects brighter than $-$16.  In contrast, the
Coma sample covered a larger area, but was not as deep in limiting magnitude.
This resulted in the Coma sample filling the luminosity range between $-$16
and $-$19.5, as well as providing a comparison sample of giant ellipticals.
As noted by other studies (Ferguson \& Binggeli 1994, Caldwell \& Bothun
1987), the dEN type galaxies are brighter, on average, than the dE types.
For the combined sample, the mean luminosity of dENs is 1.3 mags brighter
than dEs.  Since the nuclei in dENs are never brighter than 20\% of the
total light (0.2 mags, Ferguson \& Binggeli 1994), this excess is not due
to the nuclear light.  Magnitudes and colors were determined through
aperture photometry using a curve of growth analysis.  Typical errors are
0.1 in $m_{5500}$, 0.02 in $vz-yz$ and 0.01 in $bz-yz$ for the brighter
dwarfs, increasing by a factor of 3 to fainter members of our sample.

There are small differences between the two dwarf elliptical classes in
average color and these differences are in the opposite direction from
expectations based on the color-magnitude relation.  For example, the dEN
class has mean colors of $uz-yz=1.08\pm0.22$, $vz-yz=0.47\pm0.12$ and
$bz-yz=0.31\pm0.07$ (errors are the mean range of the data, not errors on
the mean) while the mean colors for the dE class are $uz-yz=1.38\pm0.29$,
$vz-yz=0.65\pm0.15$ and $bz-yz=0.29\pm0.08$.  The mean colors are similar,
although the dEN class has a statistically significant bluer mean $vz-yz$
color (lower metallicity) than the dE class.  What is significant is the
fact that the dEN sample is brighter, on average by 1.3 mags, to the dE
sample yet has a bluer mean color in contradiction to the expected higher
metallicities for higher stellar mass (i.e. redder color).  This is not a
stunning discovery since, as stated above, the color-magnitude relation
begins to break down at dwarf luminosities (Rakos \etal 2001).  For
comparison, the range in dwarf elliptical $bz-yz$ and $vz-yz$ colors
overlaps the bright ellipticals with a slightly lower mean metallicity
color.

\begin{figure*}
   \mbox{}\hfill
   \begin{minipage}[t]{0.99\linewidth}
     \centerline{\includegraphics[width=0.48\linewidth]{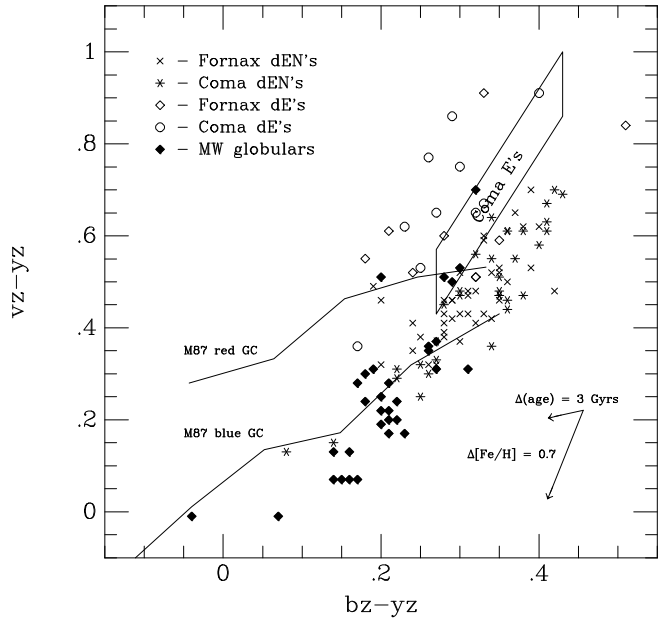}}
\caption{The multi-color diagram, $vz-yz$ versus $bz-yz$, which is the primary
diagnostic tool for understanding the star formation history in ellipticals
as $vz-yz$ is sensitive to metallicity changes and $bz-yz$ is a measure of
the stellar continuum and more sensitive to age effects.  The dwarf elliptical
data for Coma and Fornax are shown divided by morphology into nucleated dwarfs
(dEN, crossed symbols) versus non-nucleated dwarfs (dE, open symbols).  Also
shown is the region occupied by bright cluster ellipticals ($M_{5500} < -19$)
drawn from the Coma sample (95\% of all Coma ellipticals lie within the
boundaries of the indicated box).  The average colors for the red and blue
globular population in M87 from Figure 2 are also shown as solid lines along
with the data for galactic globulars (solid symbols).  For reference, the
arrows in the bottom right hand corner display the rough change in color with
age (decreases towards the blue) and metallicity for an old (13 Gyrs),
metal-rich (near solar) population from our multi-metallicity models.  }
   \end{minipage}
\end{figure*} 

Figure 4 and 5 displays the standard two color diagrams used in our previous
papers to examine the stellar populations in galaxies in distant
clusters.  Using the same notation as Rakos \etal (2001), the colors $vz-yz$ and $bz-yz$
are the metallicity and continuum indices, respectfully, due to the
sensitivity of the $vz$ filter to various metal lines near 4500\AA\ whereas
$bz$ and $yz$ are selected to cover regions relatively free of absorption or
emission features.  The $mz$ index is the standard Str\"omgren $m$ index ($mz
= (vz-bz) - (bz-yz)$), which is useful in discriminating star-forming (young
age) and non-thermal (AGN) continuum from the general shape of a galaxy's
spectral curve due to old stellar photospheres.  The crossed and open symbols
in Figure 4 represent the dENs and dEs, respectfully, in Fornax and Coma
while the marked area represents the 95\% contour of 100 bright ellipticals
in Coma (which is also in agreement with the distribution of bright ellipticals
in 20 distant clusters when corrected for passive evolution).  In addition,
the solid symbols represent 32 Milky Way globular clusters from Rakos \etal
(2001).

What is immediately obvious from these diagrams is that a majority of the
nucleated dwarf ellipticals (dEN) do not lie in a region that is a simple
extension of the normal cluster elliptical sequence extrapolated to lower
metallicities (bluer $vz-yz$ and $bz-yz$ colors).  Their average colors
are too red in $bz-yz$ for their values of $vz-yz$ (or conversely, they
are too blue in $vz-yz$ for their $bz-yz$ colors, however, we will argue
below that it is difficult to produce bluer $vz-yz$ colors for a fixed
$bz-yz$ color).  Similar behavior is found in Figure 5, the $mz$,$vz-yz$
diagram.  Nucleated dwarfs have low $mz$ indices compared to normal
cluster ellipticals, and are not an extension of the cluster elliptical
sequence.  A Kolmogorov-Smirnov (K-S) test on the distribution of dENs and
bright cluster ellipticals rules out their origin as a common population
in both diagrams at the 99.9\% level.

\begin{figure*}
   \mbox{}\hfill
   \begin{minipage}[t]{0.99\linewidth}
     \centerline{\includegraphics[width=0.48\linewidth]{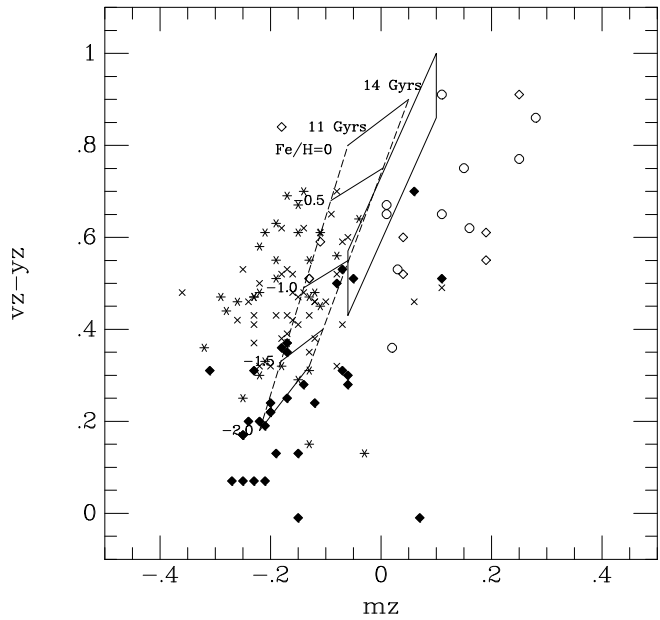}}
\caption{The $mz$ versus $vz-yz$ diagram, this color space best displays
age effects, either mean age or recent star formation.  Given the lack of
obvious star forming regions or supply of cold gas in dwarf ellipticals,
changes in this diagram are due solely to changes in the mean stellar
population age.  Symbols are the same as Figure 4, the two dotted lines
are SED models from Schulz \etal 2002 convolved to our multi-metallicity
algorithms for populations aged 11 Gyrs and 14 Gyrs.  Connecting lines of
constant mean metallicity (see Rakos \etal 2001) are also shown.  The
outlined box is the 95\% contour for bright Coma ellipticals.}
   \end{minipage}
\end{figure*} 

Also shown in Figure 4 are vectors of age (in Gyrs) and metallicity (in [Fe/H]
dex).  These vectors are for 13 Gyr populations with a mean [Fe/H] near solar,
but with a spread of internal metallicities.  Their angle and length will
vary for younger or extremely metal-poor populations, but are presented in
the Figure as a guide to the reader for the metallicity and age range expected
as normal for ellipticals in the literature.  The slope of two color plot
for dENs is shallower ($1.74\pm0.14$) than the multi-metallicity models
($4.28$) or the cluster ellipticals ($2.61\pm0.05$) and indicates that dENs
have a narrow spread of internal metallicities.  Even extremely metal-rich
cluster ellipticals must have a significant population of metal-poor stars
to explain the details of their color-metallicity relationship (Rakos \etal
2001).  This reflects in a particular slope in two color diagrams such as
Figure 4, a slope that is not reproduced for the dENs. 

An interesting point to note is that the dENs in Figures 4 and 5 display
more in common with globular clusters than cluster ellipticals with respect
to their integrated colors.  Normal ellipticals are not simple stellar
populations, i.e. one age and one metallicity, so it is not surprising to
find that they are not a linear extension of the globular cluster sequence to
redder colors (i.e.  higher metallicities) as can be seen in Figure 4.  In
fact, in Rakos \etal (2001), we were able to model the colors of bright
cluster ellipticals in a non-linear fashion, using multi-metallicity
models based on globular clusters and SSP models.  We might naively expect
the colors of low mass systems to approach the colors of SSP's, like
globular clusters, due to a much more limited range of metallicities in
the underlying stellar population.  However, Figure 4 indicates that dENs
appear to be an exact extrapolation of the globular cluster sequence,
which are a simple stellar population (i.e. composed of stars of a single
metallicity) and suggests that the spread in the internal metallicities of
dENs is very small.

In contrast to the dENs, the non-nucleated dwarf ellipticals (dE) lie parallel
to the bright elliptical sequence but to the blueward side of the mean $bz-yz$
color for cluster ellipticals.  This division is also obvious in the $mz$,$vz-yz$
diagram where all the normal and dE ellipticals have $mz$ indices greater
than $-$0.1, but the globulars and dENs are primarily below $-$0.1 regardless
of $vz-yz$ color.  This same difference is seen in all four narrow band colors
($uv-vz$, $uz-yz$, $vz-yz$ and $bz-yz$) and clearly divides the morphological
populations of dwarf ellipticals into two groups based solely on the color
of the underlying stellar populations.  In this sense, we confirm the results
on the early-type galaxies in Perseus by Conselice, Gallagher \& Wyse (2002),
that there appear to be two separate elliptical populations in rich clusters;
a bright one, containing normal cluster ellipticals and a faint one containing
low mass systems.

It is also important to note that this color difference between dEs and dENs
cannot be due to the color of the nucleus in dEN class objects since surface
photometry (Caldwell \& Bothun 1987) has shown that the nucleus contributes
only between 10 to 15\% of the total light of the galaxy.  This would require
that the nucleus have a mean $vz-yz$ color of $-$0.08 to change the integrated
colors of a normal elliptical into the typical color of a dEN.  There is no
indication of an extremely blue nucleus in any of our inner apertures, nor
was there any sign of color gradients in the Caldwell \& Bothun photometry
of Fornax dENs.  Additionally, a sharp change in $vz-yz$ color will produce
a corresponding blueward shift in $bz-yz$ of 0.16 mags which makes it impossible
to explain a majority of the red dENs as blue contaminated normal elliptical
colors.

Given the detection of two color populations of globular clusters in other
galaxies (Burgarella, Kissler-Patig \& Buat 2001), it is worthwhile
comparing the colors of the dwarf ellipticals presented herein with those
simpler stellar populations of globular clusters.  As discussed in \S2.1,
the blue population of globulars in M87 is very similar to the Milky Way
globulars, a uniform sample of old, metal-poor systems.  The sequence of
color for the blue population matches well with the sequence of
metallicity found in our own Galaxy's globulars.  In addition, as can be
seen in Figure 4, the blue population of globulars in M87, galactic
globulars and the nucleated dwarfs in Fornax and Coma (dENs) all form the
same sequence of objects whose range of colors are due solely to
metallicity.  In contrast, the red population of globulars in M87 lie to
the upper left of the blue population similar to the relationship of the
colors of non-nucleated dwarfs (dEs) and normal cluster ellipticals to
dENs.  The red population does not reach the $vz-yz$ nor $bz-yz$ values of
cluster ellipticals, presumably they do not reach the near-solar
metallicities of ellipticals.

Figure 6 displays the behavior of the dwarf galaxies in the near-UV color,
$uz-yz$ versus continuum color $bz-yz$.  As before, the dEN galaxies follow
the galactic globular colors as well as the M87 blue population of globulars.
While there is some intermixing of dE types, a majority lie to the redward
side of $uz-yz=1.2$ and, again, follow the red population of globulars.
The dashed line is the track of 13 Gyrs models from the G\"ottingen SSP
models (Schulz \etal 2002), convolved to our metallicity model.  As found
in our distant cluster work (Rakos \& Schombert 1995), $uz-yz$ is poorly
modeled by any group's SED's as it cuts laterally across the color plot.
While the dEN data lie too blue in $uz-yz$ as compared to the models, it
is not possible to determine if this is due to the lack of some blue
component to the UV data in the models (e.g. blue HB stars) or some very
young population in dwarf ellipticals.

\section{DISCUSSION}

The separation of dwarf ellipticals into nucleated and non-nucleated leads
to a surprising dichotomy with respect to their mean colors.  The trend can
be summarized in Figure 4 where the nucleated dwarfs (dEN class) are redder
in $bz-yz$ and bluer in $vz-yz$ than normal cluster ellipticals (also lower
$mz$ indices and bluer $uz-yz$ colors).  In contrast, non-nucleated dwarfs
(dE class) lie in the opposite direction with respect to normal cluster
ellipticals (i.e., bluer in $bz-yz$, see Figure 5).  The normal cluster
ellipticals are a color population that divides the two dwarf classes, although
they overlap with the dE galaxies.

With respect to star formation history, almost any portion of the two color
diagram can be reached by some combination of metallicity, recent star
formation or extinction.  However, ellipticals have several limitations in
their recent star formation histories that can be further constrained by
comparison with other filters as well as spectroscopic and 21-cm studies.
Through numerous studies of the gas and spectral properties of ellipticals
(Trager \etal 2000), it has been demonstrated that recent star formation for
cluster ellipticals is extremely limited over the past 5 Gyrs, in agreement
with their smooth, lacking in star-forming knots, morphological appearance.
The lack of far-IR emission from warm dust (Knapp \etal 1989) also eliminates
reddening by extinction as a major contribution.

A recent star formation history that is quiescent allows us,
qualitatively, to consider the evolution of ellipticals through a singular
combination of age and metallicity, assuming that the age of the embedded
populations is more than 5 Gyrs.  In order to draw quantitative
conclusions on the meaning of the color differences outlined above, one
must consult galaxy evolution models that blend the populations with
various metallicities in a galaxy with their calculated variation with
time.  Thus, our conclusions in this section are mildly model-dependent,
and the reader is cautioned to fully explore the limitations of SED models
before drawing precise numbers from our data (such as absolute age in
Gyrs).

\begin{figure*}
   \mbox{}\hfill
   \begin{minipage}[t]{0.99\linewidth}
     \centerline{\includegraphics[width=0.48\linewidth]{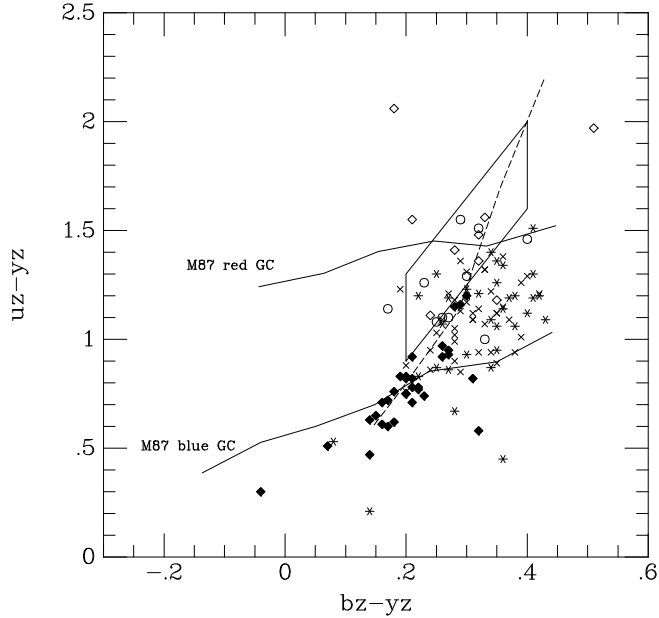}}
\caption{The near-UV color ($uz-yz$) versus continuum color ($bz-yz$).
Symbols are the same as previous figures.  Also shown are the M87 globular
cluster tracks and the 13 Gyrs G\"ottingen SSP model (Schulz \etal 2002, dashed
line).  Outlined box is the 95\% contour for bright Coma ellipticals.}
   \end{minipage}
\end{figure*} 

As discussed in Rakos \etal (2001) and Odell, Rakos \& Schombert (2002),
relative success in explaining the color-magnitude and the metallicity-color
relation in ellipticals has been achieved through the use of a simple,
multi-metallicity model where mean age and mean metallicity are the only free
parameters.  These models are constructed using the SSP SED's available in
the literature, primarily Schulz \etal (2002), where an estimate of the
distribution of metallicity is made based on a simple chemical evolution
scenario (see Kodama \& Arimoto 1997).  In brief, our multi-metallicity models
start with the concept that a galaxy is composed of a smooth continuum of
stars of varying metallicities. It is obvious from inspection of the stellar
populations in our own Galaxy, that the colors of a composite system, such as
an elliptical, cannot be reproduced by using a single stellar population
with the mean metallicity of the galaxy.  A real galaxy is composed of many
stars with a range of metallicities that sum to the luminosity weighted
average.  From the physics of stellar atmospheres, we know that each group
of stars with a particular metallicity will contribute very different amounts
of light into the various filters.  In order to correct for this effect, we
elected to mix different SSP's using a simple distribution of metallicity
that conforms to our expectations based on observations of our own Galaxy
and chemical evolution models (Kodama \& Arimoto 1997).  First, the underlying
metallicity distribution is assumed to be Lorentzian in shape, a long tail to
low metallicities and a sharp cutoff at high metallicities.  A range of galaxy
metallicities is constructed by holding the lowest metal population fixed
([Fe/H]=$-2.5$) and sliding the upper cutoff from high to low.  The Str\"omgren
colors for each metallicity bin are determined by empirical calibration to
globular clusters ([Fe/H]$<-1$) and spectrophotometric SSP models ([Fe/H]$>-1$,
Schulz \etal 2002).  These simple models do an excellent job of reproducing
the color-metallicity relationship ($vz-yz$ versus [Fe/H]) and, assuming a
$M/L$ ratio of 5 for early-type galaxies, of mapping the color-magnitude
relation into the observed mass-metallicity relation (Odell, Rakos \&
Schombert 2002)

As an example of this process, the 11 and 14 Gyr multi-metallicity models
are shown in Figure 5.  For variety, tracks in Figure 5
are based on the G\"ottingen SSP models (Schulz \etal 2002) rather than the
Bruzual \& Charlot models as in our last few papers, but the differences are
minor and use of either set of models does not change our basic interpretation.
The models predict a broad range of internal metallicities for high mass
ellipticals, decreasing to a narrower range for lower mass systems (i.e.
bluer colors).  We note that all the models fail to reproduce the upturn in
$vz-yz$ color for lowest metallicity galactic globulars in the $vz-yz$,$bz-yz$
diagram.  This is probably due to a failure in the SED models to correctly
track the effect of hot horizontal branch stars (see Leonardi \& Rose 2003).
Calibration to a metallicity scale is achieved through various comparisons
with spectroscopy work (Salaris \& Weiss 1998), but age calibration is highly
uncertain due to a lack of young, metal-poor objects around our Galaxy.

Considering the dEN ellipticals first, as a class they lie in a region of
the multi-color diagrams that are redder in continuum colors while maintaining
the same slope in $bz-yz$,$vz-yz$ as given by a metallicity effect.  When
compared to the multi-metallicity models, the dEN objects have all the
indications of a population that has the same color-metallicity relation as
cluster ellipticals (although for lower total metallicities) but is older by
approximately 2 to 3 Gyrs (redder continuum colors).  This would make dENs
some of the oldest stellar objects in the local Universe, and the fact that
the color sequence of galactic globulars, and the blue population in M87,
track precisely into the dEN colors would seem to confirm their old age.

It is well established that the colors of bright, early-type galaxies in
clusters are dominated by metallicity effects rather than age effects
(Kuntschner \& Davies 1998, Vazdekis \etal 2001).  This, apparently, is also
true of dwarf ellipticals, in that the spread of dEN colors tracks along a
metallicity vector, not the age vector (see Figure 4).  The range of
metallicities for the dEN galaxies is similar to that found in previous
broadband and spectroscopic studies, ranging from $-$1.2 to $-$0.4 [Fe/H].
Cluster ellipticals range from $-$0.7 to $+$0.2, corrections for age would
only adjust these values by 0.1 dex (in the direction of making the cluster
ellipticals more metal-rich).  The metallicity values of the lowest [Fe/H]
dENs merge with the more metal-rich galactic globulars (e.g. NGC 204 and
Pal 12).   Interestingly, none of the cluster ellipticals or dE class galaxies
obtain the lowest [Fe/H] values of the dENs, again supporting the idea that
dENs are older than dEs or bright cluster ellipticals. In other words, the
dENs have had a shorter phase of initial star formation and have undergone
less chemical evolution.

The interpretation for the dE galaxies is opposite to that for dENs, 
such that we have a sequence which is bluer in continuum colors, indicating a
younger mean age as compared to dENs.   The mean color of dEs is also
slightly bluer than normal cluster ellipticals, although the overlap
between dEs and normal cluster ellipticals is greater than the
overlap for dENs, making the dE population appear as more of a
continuation of the normal elliptical track with slightly bluer continuum
colors.  As before, the model driven interpretation of these color
differences is that dEs, as a group, are slightly younger than cluster
ellipticals by about 1 to 2 Gyrs.  This type of behavior, with respect to
mean age, has been seen in spectroscopic studies.  For example, Poggianti
\etal (2001) completed a spectroscopic study of Coma down to $M_B=-14$.
Assigning metallicity and age through various Fe and Balmer line indices,
they find a bimodal distribution of galaxy ages at low luminosity.  In
fact, their Figure 7 displays a clear division at faint luminosities into
galaxies with ages greater than 9 Gyrs (our dENs) and those with ages
younger than 5 Gyrs (our dEs).  Their age scale is not sensitive enough to
compare the ages of old objects and, thus, lack evidence that the old
dwarf galaxies are older than the oldest bright ellipticals as we have
shown above.

This new perspective on the colors of ellipticals (both bright and dwarf)
eliminates a great deal of the confusion in the interpretation of multi-color
diagrams with respect to the underlying stellar populations in ellipticals.
For example, the range of mean ages found in Fornax ellipticals increased
with decreasing luminosity, but not in a linear or systematic manner (Rakos
\etal 2001).  We now understand that most of this scatter is due to including
both dE and dENs in the analysis and the fact that the dwarf ellipticals have
no correlation between metallicity color and luminosity.  This also drives
the CMR in $bz-yz$ to redder colors for lower luminosities (as noted in Odell,
Rakos \& Schombert 2002) producing what appears to be an age effect for all
ellipticals when, in fact, the inclusion of old dENs into the sample is the
primary reason as they have mean ages that are older than cluster ellipticals.

The dichotomy by color between dEs and dENs would also explain a great
deal of divergence in the literature concerning the ages of ellipticals.
Recent spectroscopic work by Trager \etal (2000), Kuntschner (2000) and
Caldwell \etal (2003) have all converged an age for cluster ellipticals
which is old (greater than 10 Gyrs).  However, Caldwell \etal also find
that, for low velocity dispersion galaxies (i.e. low mass), the mean age
is about 4 to 6 Gyrs younger than bright field or cluster ellipticals.
While there are very few dEs in the Caldwell \etal sample, morphologically
the sample is similar to low luminosity E and dEs herein.  Their
determination of a low mean age from spectroscopic lines agrees well with
our global continuum color determination, as their sample is dominated by
low luminosity E, S0's and dEs with very few dENs.

In summary, our current narrow band dataset extends from globular clusters
to the bright cluster elliptical in rich clusters.  When taken as a whole,
there is a broad range of metallicities and ages that can explain any particular
galaxy's color.  However, when subdivided by galaxy type, a very different
picture of galaxy formation and evolution appears.  Two separate and distinct
populations are indicated by the data.  One population, composed of galactic
globulars, the blue population of extragalactic globulars and nucleated dwarf
ellipticals (dEN), displays the clear trend of objects composed of old stars
(at least 13 Gyrs based on Milky Way CM diagrams) with increasing metallicities
similar to standard SSP models.  As with globulars, there is no mass-metallicity
relation for the dENs, i.e. the mean [Fe/H] is independent of luminosity,
which implies that they are not self-enriched like bright cluster ellipticals.
And, unlike cluster ellipticals, the slope in the two color diagrams indicate that
the colors of dENs have a very narrow spread of internal metallicity, which
explains their similarity in color space with globulars.  An old age for the
dEN class is also consistent with their cluster spatial distribution, which
is much more compact than that of dEs (Conselice, Gallagher \& Wyse 2001,
Ferguson \& Sandage 1989) where hierarchical clustering models predict the
oldest galaxies should be located (White \& Springel 2000).

While the dENs appear to be old objects, we agree with the conclusions of
Conselice, Gallagher \& Wyse (2003), drawn from a study of low mass galaxies
in the Perseus cluster, that there exists a second population in clusters
which cannot be primordial.  The second population is composed of the red,
metal-rich globulars, non-nucleated dwarf ellipticals (dE)
and bright cluster ellipticals.  The galaxies in this group display a clear
mass-metallicity relation, but one that requires the elliptical's stellar
populations be a composite of many groups of stars with varying
metallicities.  Regardless of the calibration for age from the SED models,
the cluster ellipticals, dEs and the red population of globulars in M87
are located in a region of the multi-color diagram occupied by bluer
stellar populations.  This leaves us to conclude that they are younger, as
a group, than the blue population of galactic globulars and dENs.
Comparison to models places this age difference as between 3 to 4 Gyrs.

The implications of two early-type populations in clusters with respect to
galaxy formation and evolution scenarios is varied.  While all indications
are that ellipticals are some the oldest objects in clusters, the data
presented here suggests that only the blue galactic globulars and dENs are
the primordial objects.  Other cluster ellipticals and dEs have either a
later formation time or an initial star formation burst that was prolonged
which produced a mean younger age as given by integrated colors.  Our data are 
also consistent with the idea that dEs are remnants of more massive galaxies
where a majority of their mass has been stripped away.  However, this scenario
is inconsistent with their structural properties, i.e. they do not have
central densities similar to bright ellipticals (Guzman \etal 2003) and,
thus, we reject it for the following discussion.

The connection between the red globular population and ellipticals would
argue for a late formation epoch for ellipticals, in that the red cluster's
higher metallicities are not due to internal chemical evolution but produced
by their birth in higher metallicity clouds, enriched by previous star
formation (presumingly, the epoch that produced the blue globular population).
However, the color-metallicity diagram for ellipticals demonstrates that the
global characteristics of ellipticals cannot be reproduced by a simple,
single metallicity population (Rakos \etal 2001).  Their internal stellar
population must include a significant metal-poor component and, thus, an
extended initial star formation era would best explain their younger age
compared to primordial globulars.  Yet, this initial epoch of star formation
cannot have been too prolonged or else extensive color gradients would form in
ellipticals.  In addition, a prolonged burst would be inconsistent with bright
elliptical $\alpha$/Fe ratios (Thomas \& Kauffmann 1999).  It remains an open
question for the galaxy modeling community whether the estimated 3 to 4 Gyrs
age difference between cluster and dwarf ellipticals can be reproduced with
small metallicity gradients without resorting to a later epoch of formation
than dENs and metal-poor globulars.

The fact that the colors of the red population of M87 globulars tracks into
the color sequence of cluster ellipticals and dE types implies that the
metal-rich globulars formed in the enriched gas produced by earlier star
formation.  Comparison to model tracks dates this era as approximately 2 Gyrs
after the formation of the blue, metal-poor globulars.  Given the rapid
enrichment of the environment near ellipticals, it appears that the blue
population of globulars in galaxies, such as M87, is acquired later through
mergers of whole galaxy systems or cannibalism of stray globulars in the
cluster environment.  This would seem contrary to the arguments that the
metal-poor globular population must have formed with its parent galaxy since
it is homogeneous.  However, the metal-rich population matches the color of
the underlying galaxy envelope and the mean metallicity of the red population
of globulars is correlated with the mass of the elliptical (Larsen \etal
2001).  Thus, the homogeneous nature and narrow range in metallicity for the
metal-poor globulars may be signaling an origin that is common to all galaxies,
such as accretion from a common intracluster pool.

Our new data on ages of dEs and dENs provides an opportunity to revisit the
entire issue on the ages of passively evolving systems, such as ellipticals,
in clusters.  Most spectroscopic studies have found that bright ellipticals
in rich clusters are fairly coeval in their stellar populations (Gonzalez
1993, Trager \etal 2000, Kuntschner 2000) with very little variation in age
with luminosity.  This would confirm that the uniform slope in the CMR is
due to solely a metallicity effect.  Our interpretation of Figure 4, the
$vz-yz$,$bz-yz$ diagram, is that the bright cluster ellipticals in Coma have
a two-color correlation that is similar in slope to a single age population
varying only in metallicity.  There is a slight shift in color, as mentioned
in Odell, Rakos \& Schombert (2002), such that low luminosity ellipticals
appear to be younger than their more massive cousins (i.e. bluer continuum
colors).  However, we also note that there is a variation in $\alpha$ element
abundances with luminosity (see Thomas, Maraston \& Bender 2002, Trager \etal
2000, Rakos \etal 2001) such that bright ellipticals have $\alpha$/Fe values
greater than dwarf ellipticals.  The $\alpha$/Fe index is a measure of the
duration of the burst of initial star formation where low $\alpha$/Fe signal
prolonged bursts giving time for Type I SN to increase the abundance of Fe.
Under this condition, massive ellipticals have shorter durations of star
formation compared to low luminosity ellipticals.  And longer initial star
formation duration would reflect in a younger mean age, as is confirmed in
our data for dEs.

A dissenting view concerning the age of ellipticals is offered by Terlevich
\& Forbes (2002) who find that brighter ellipticals (i.e. more metal-rich)
have younger mean ages.  This, of course, would make sense in a scenario
where metal enrichment is due to a later formation epoch and the gas clouds
are overabundant from previous star formation.  However, the required ages
for massive ellipticals would be less than 5 Gyrs, which is in disagreement
with observations of passive evolution for high redshift clusters (Rakos \&
Schombert 1995, Stanford, Eisenhardt \& Dickinson 1998).

\section{CONCLUSIONS}

This study presents Str\"omgren narrow band photometry of dwarf ellipticals
in the Fornax and Coma clusters, combining samples from previous work with
some new objects and colors (e.g. $uz-yz$).  Our findings can be divided
into three parts: pure observations, multi-color analysis and model
dependent interpretation, listed herein in descending order of confidence:

\begin{itemize}

\item{} Several basic characteristics of nucleated dwarf ellipticals (dEN)
and non-nucleated dwarf ellipticals (dE), that were discovered by previous
broadband studies (see Caldwell \& Bothun 1987), are confirmed.  In terms of
total numbers, dENs are more numerous than dEs and brighter, on average, by
1.3 mags.  The central nucleus, while obvious in physical morphology and
surface photometry, rarely contributes more than 10 to 15\% of the total
light (mass) of the galaxy, typical the size and luminosity of a supermassive
star cluster.

\item{} In our multi-color diagrams ($vz-yz$,$bz-yz$, $mz$,$vz-yz$ and
$uz-yz$,$bz-yz$), the population of dENs are distinct from the colors of
dEs or normal cluster ellipticals.  In general, the dENs are redder in
continuum color ($bz-yz$) and bluer in metallicity color ($vz-yz$).  Normal
dwarfs (dE) are slightly bluer in continuum colors with bright cluster
ellipticals forming a color sequence intermediary between the two types of
dwarfs.  This confirms the results of Conselice, Gallagher \& Wyse (2001),
that there are two populations of early-type galaxies in rich clusters.

\item{} In comparison with globular cluster colors, dENs appear to be a linear
extrapolation of the galactic globular colors and the colors of the blue
populations of globulars in other galaxies (i.e. M87).  Cluster ellipticals
and dEs track the red, metal-rich population of extragalactic globulars. 

\item{} Comparing to multi-metallicity models, and assuming that the populations
in ellipticals can be described by a set of stars with a smooth distribution
of metallicity and with ages greater than 8 Gyrs, then dENs are 3 to 4 Gyrs
older than bright cluster ellipticals and about 5 Gyrs older than dE type
galaxies.  Of the elliptical class, dENs span the lowest metallicities
reaching levels that are 0.7 dex below the metal-poorest bright cluster elliptical.
This would support their old age based on the age-metallicity relation for
ellipticals from Terlevich \& Forbes (2002).

\item{} When placing the various ellipticals and globular systems in order
of their relative ages, it appears that only the blue globulars and dENs have
solely primordial stellar populations.  Cluster ellipticals and dEs either
have later formation epochs or a prolonged burst of star formation compared
to dENs.  However, a lengthy phase of initial star formation does not support
the lack of steep color gradients in ellipticals.

\end{itemize}

The data presented herein does not to support, or refute, any particular galaxy
formation scenario.  The most popular scenario, under the CDM framework, is
one in which galaxies form through a hierarchical sequence of mergers (e.g.
Davis \etal 1985).  This would predict that, in rich environments, high mass
galaxies assembled last and, thus, would have the youngest mean age but that
there would be a large scatter in age per mass bin.  And, of course, dwarf galaxies
would be the initial seeds to galaxy construction and would have the oldest
mean ages.  The problem with testing these predictions on cluster ellipticals
is that any later star formation, or accretion of a star-forming object such
as a spiral, between the formation epoch and the present will produce a young,
luminosity weighted, mean age in our narrow band color system even if a
majority of the galaxy's stellar population is old.  Low metallicity and old
age for the dENs implies a primordial origin and if the younger age and higher
metallicity dEs have the same formation process then they must have a different
formation epoch.  Our current data does support the conjecture that {\it
some} dwarf galaxies are primordial (i.e.  dENs).  But, not all cluster dwarf
ellipticals are old, thus, the formation of young dEs may involve stripping
of high mass ellipticals as proposed by Conselice \etal (2003).

Lastly, we consider these stellar population results for the whole class
of ellipticals in light of their structural and kinematic properties.  The
notion that ellipticals divide into bright ellipticals and dwarf
ellipticals by structure (power-law versus exponential profiles) dates
back to ground-based surface photometry in the 1980's (see Kormendy
1985).  Thus, there was some expectation that dwarfs and bright
ellipticals would have varying star formation histories or even completely
different formation processes.  However, recent work with HST imaging
(Graham \& Guzman 2003) has demonstrated that the two morphological types
are indeed similar in underlying structure.  In this sense, the connection
between dE and normal cluster ellipticals becomes one of decreasing mass
and metallicity, with some evidence that lower mass ellipticals are
slightly younger than more massive cluster ellipticals (Caldwell \etal
2003).  The connection between globular clusters, dENs and the new
`ultracompact' class of galaxies (Phillipps \etal 2001) remains an intriguing
line of future inquiry.

\acknowledgements The authors wish to thank NOAO and Steward Observatory for
granting telescope time for this project.  One of us (K. Rakos) gratefully
acknowledges the financial support from the Austrian Fonds zur Foerderung
der wissenschaftlichen Forschung.  Special thanks to Andres Jordan for sharing
his HST data on the M87 globulars and to Chris Conselice for his suggestions.
We acknowledge the support of NASA under grants HST-AR-08758 as well as tools
developed under AIRSP NAG5-11300.


\begin{references}

\reference{} Arimoto, N. \& Yoshii, Y. 1987, \aap, 173, 23
\reference{} Bender, R., Burstein, D. \& Faber, S. 1993, \apj, 411, 153
\reference{} Bothun, G. \& Mould, J. 1988, \apj, 324, 123
\reference{} Bothun, G., Mould, J., Wirth, A. \& Caldwell, N. 1985, \aj, 90, 697
\reference{} Bruzual, A. \& Charlot, S. 1993, \apj, 405, 538
\reference{} Burgarella, D., Kissler-Patig, M. \& Buat, V. 2001, \aj, 121, 2647
\reference{} Caldwell, N. \& Bothun, G. 1987, \aj, 94, 1126
\reference{} Caldwell, N., Rose, J. \& Concannon, K. 2003, astro-ph/0303345
\reference{} Cohen, J., Blakeslee, J. \& Ryzhov, A. 1998, \apj, 496, 808
\reference{} Conselice, C., Gallagher, J. \& Wyse, R. 2003, \aj, 125, 66
\reference{} Conselice, C., Gallagher, J. \& Wyse, R. 2001, \apj, 559, 791
\reference{} Conselice, C., Gallagher, J. \& Wyse, R. 2002, \aj, 123, 2246
\reference{} Conselice, C., Gallagher, J. \& Wyse, R. 2003, \aj, 125, 66
\reference{} Davis, M., Efstathiou, G., Frenk, C. \& White, S. 1985, \apj, 292, 371
\reference{} Eggen, O., Lynden-Bell, D. \& Sandage, A. 1962, \apj, 136, 748
\reference{} Evans, R., Davies, J. \& Phillipps, S. 1990, \mnras, 245, 164
\reference{} Ferguson, H. 1989, \aj, 98, 367
\reference{} Ferguson, H. \& Binggeli, B. 1994, \araa, 6, 67
\reference{} Ferguson, H. \& Sandage, A. 1989, \apj, 346, L53
\reference{} Gebhardt, K. \& Kissler-Patig, M. 1999, \aj, 118, 1526
\reference{} Gerola, H., Carnevali, P. \& Salpeter, E. 1983, \apj, 268, L75
\reference{} Gonzalez, J. 1993, Ph.D. Thesis, UCSC
\reference{} Graham, A. \& Guzman, R. 2003, \aj, 125, 2936
\reference{} Guzman, R., Graham, A., Matkovic, A. \& Vass, I. 2003, astro-ph/0303390
\reference{} Held, E. \& Mould, J. 1994, \aj, 107, 1307
\reference{} Jordan, A., Cote, P., West, M. \& Marzke, R. 2002, \apj, 576, L113
\reference{} Kauffmann, G. \& Charlot, S. 1998, \mnras, 294, 705
\reference{} Kissler-Patig, M. 2001, {\it Extragalactic Star Clusters: IAU Symposium Series} 207, 333
\reference{} Knapp, G., Guhathakurta, P., Kim, D. \& Jura, M. 1989, \apjs, 70, 329
\reference{} Kodama, T. \& Arimoto, N.1997, \aap, 320, 41
\reference{} Kormendy, J. 1985, \apj, 295, 73
\reference{} Kundu, A., Whitmore, B., Sparks, W., Macchetto, F., Zepf, S. \& Ashman, K. 1999, \apj, 513, 733
\reference{} Kuntschner, H. 2000, \mnras, 315, 184
\reference{} Kuntschner, H. \& Davies, R. 1998, \mnras, 295, L29
\reference{} Larsen, S., Brodie, J., Huchra, J., Forbes, D. \& Grillmair, C. 2001, \aj, 121, 2974
\reference{} Larson, R. 1974, \mnras, 166, 585
\reference{} Leonardi, A. \& Rose, J. 2003, astro-ph/0306358
\reference{} Madore, B., Freedman, W., Silbermann, N., Harding, P., Huchra, J., Mould, J., Graham, J., Ferrarese,
L., Gibson, B., Han, M., Hoessel, J., Hughes, S., Illingworth, G., Phelps, R., Sakai, S. \& Stetson, P. 1999,
\apj, 515, 29
\reference{} Odell, A., Schombert, J. \& Rakos, K. 2002, \aj, 124, 3061
\reference{} Oh, K. \& Lin, D. 2000, \apj, 543, 620
\reference{} Phillipps, S., Drinkwater, M., Gregg, M. \& Jones, J. 2001, \apj, 560, 201
\reference{} Poggianti, B., Bridges, T., Mobasher, B., Carter, D., Doi, M., Kashikawa, N., Komiyama, Y., Okamura, S. \& Sekiguchi, M. 2001, \apj, 562, 689
\reference{} Puzia, T., Kissler-Patig, M., Brodie, J. \& Huchra, J. 1999, \aj, 118, 2734
\reference{} Puzia, T., Zepf, S., Kissler-Patig, M., Hilker, M., Minniti, D. \&
Goudfrooij, P. 2002, \aap, 391, 453
\reference{} Rakos, K. \& Schombert, J. 1995, \apj, 439, 47
\reference{} Rakos, K., Schombert, J., Maitzen, H., Prugovecki, S. \& Odell, A. 2001, \aj, 121, 1974
\reference{} Salaris, M. \& Weiss, A. 1998, \aap, 335, 943
\reference{} Schulz, J., Fritze-v, Alvensleben,, Moumller, C. \& Fricke, K. 2002, \aap, 398,
89
\reference{} Somerville, R. \& Primack, J. 1999, \mnras, 310, 1087
\reference{} Stanford, S., Eisenhardt, P. \& Dickinson, M. 1998, \apj, 492, 461
\reference{} Steindling, S., Brosch, N. \& Rakos, K. 2001, \apjs, 132, 19
\reference{} Terlevich, A. \& Forbes, D. 2002, \mnras, 330, 547
\reference{} Terlevich, A., Kuntschner, H., Bower, R., Caldwell, N. \& Sharples, R. 1999, \mnras, 310, 445
\reference{} Thomas, D. \& Kauffmann, G. 1999, {\it Spectrophotometric Dating of Stars and
Galaxies, ASP Conference Proceedings}, Vol. 192, 261
\reference{} Thomas, D., Maraston, C. \& Bender, R. 2002,  {\it Astronomy with Large
Telescopes from Ground and Space, Reviews in Modern Astronomy}, Vol. 15, 219
\reference{} Tonry, J., Dressler, A., Blakeslee, J., Ajhar, E., Fletcher, A., Luppino, G., Metzger, M. \& Moore, C. 2001, \apj, 546, 681
\reference{} Trager, S., Faber, S., Worthey, G. \& Gonzalez, J. 2000, \aj, 120, 165
\reference{} Vazdekis, A., Kuntschner, H., Davies, R., Arimoto, N., Nakamura, O. \& Peletier, R. 2001, \apj, 551, L127
\reference{} White, S. \& Frenk, C. 1991, \apj, 379, 52
\reference{} White, S. \& Springel, V. 2000, {\it The First Stars. Proceedings of the MPA/ESO
Workshop}, 327

\end{references}
\end{document}